\documentclass{kluwer}    
\usepackage[dvips]{graphicx}

\begin{document}                                                                                   
\begin{article}
\begin{opening}         
\title{The Interplay of Cluster and Galaxy Evolution}
\author{Robert C. \surname{Nichol}, Christopher J. Miller \&  Tomotsugu Goto}  
\runningauthor{Nichol, Miller \& Goto}
\runningtitle{Interplay of Cluster and Galaxy Evolution}
\institute{Dept. of Physics, Carnegie Mellon University, 5000 Forbes Ave., Pittsburgh, PA-15232, USA (nichol@cmu.edu, chrism@cmu.edu, tomo@cmu.edu)}

\begin{abstract}
We review here the interplay of cluster and galaxy evolution. As a
case study, we consider the Butcher--Oemler effect and propose that it
is the result of the changing rate of cluster merger events in a
hierarchical universe. This case study highlights the need for new
catalogs of clusters and groups that possess quantified
morphologies. We present such a sample here, namely the Sloan Digital
Sky Survey (SDSS) C4 Catalog, which has been objectively--selected
from the SDSS spectroscopic galaxy sample. We outline here the C4
algorithm and present first results based on the SDSS Early Data
Release, including an X--ray luminosity--velocity dispersion (${\rm
L_x}$--$\sigma_v$) scaling relationship (as a function of cluster
morphology), and the {\it density--SFR} relation of galaxies within C4
clusters (Gomez et al. 2003). We also discuss the merger of Coma and
the NGC4839 group, and its effect on the galaxy populations in these
systems. We finish with a brief discussion of a new sample of
H$\delta$--selected galaxies ({\it i.e.}, k+a, post--starburst
galaxies) obtained from the SDSS spectroscopic survey.
\end{abstract}
\end{opening}           

In this review, we investigate the interplay of cluster and galaxy
evolution. We do so via two case studies: First, the relation between
the Butcher--Oemler effect and the changing rate of cluster merger
events in a hierarchical universe. Secondly, a study of the
distribution of post--starburst galaxies in, the around, the Coma
Cluster of Galaxies.

\begin{figure}[t]
\centerline{\includegraphics[width=24pc]{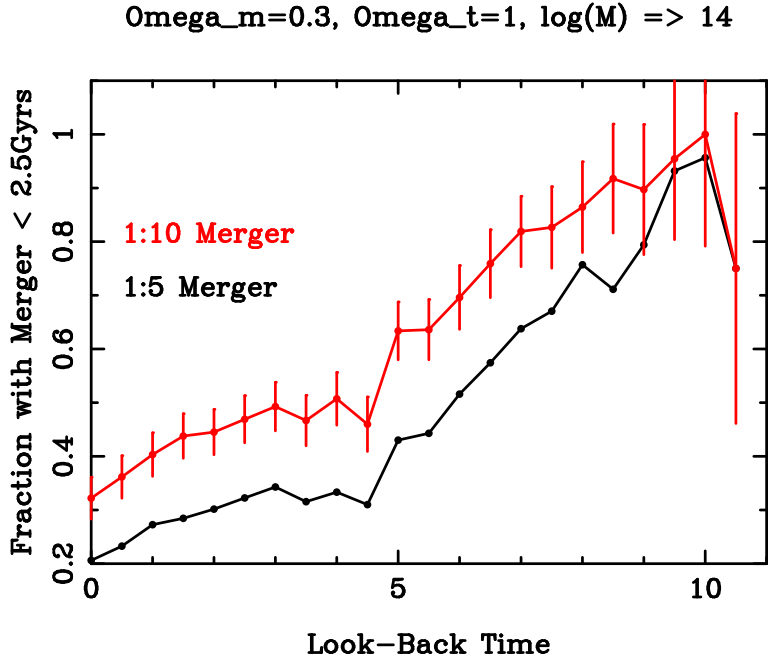}}
\caption[merger]{The fraction of $M\ge10^{14}\,M_{\odot}$ clusters
that have experienced a major merger within the last 2.5Gyrs versus
look-back time.  The red points are for a 10:1 merger, while the black
points are for a 5:1. These data are kindly provided by Joanne Cohn
based on simulations similar to the work presented in Cohn, Bagla \&
White (2000). The simulations assume $\Omega_{m}=0.3$,
$\Omega_{\Lambda}=0.7$ and $\sigma_8=0.9$. The error bars are
$\sqrt{N}$ of clusters detected in the simulations. For clarity, they
are not shown for the 5:1 simulations.}
\label{merger}
\end{figure}

\section{Cluster Mergers and the Butcher--Oemler Effect}

In a hierarchical universe, clusters of galaxies are formed through
the merger of small clusters and groups of galaxies ({\it e.g.},
Dubinski 1998). Such merger events have a significant effect on the
physical nature of the clusters involved, as such events can release a
total kinetic energy of $\sim10^{63}$ ergs\footnote{By comparison,
supernovae and gamma--ray bursts release $\sim 10^{54}$ ergs of
energy}. This energy drives shocks through the intra-cluster medium
(ICM), heating the gas and generating turbulence. Detailed simulations
of major cluster mergers ({\it e.g.}, mass ratio of 3:1) show that the
effect of such events can persist for over 2 Gigayears, with
large-scale eddies running through the ICM, up to several hundreds of
kpcs in size.  Even after a Hubble time, these motions persist as
subsonic turbulence in the cluster cores, providing 5 to 10\% of the
pressure support of the ICM against gravity (see Ricker \& Sarazin
2001; Kempner, Sarazin \& Ricker 2002; Roettiger, Burns \& Loken 1996;
Ritchie \& Thomas 2001; Schindler et al. 2001).

In Figure \ref{merger}, we present the expected fraction of clusters
with $M\ge10^{14}\,M_{\odot}$ ($\sim {\rm L_x(0.5-2.0 keV)} > 10^{43}\,{\rm
erg/s}$) that have experienced a major merger in the last 2.5
Gigayears, as a function of look-back time. This figure demonstrates
that as we push to higher redshift, a majority of clusters should show
significant evidence of a recent merger which may have severely affected
their physical state (see Mathiesen \& Evrard 2001). Figure
\ref{merger} is in qualitative agreement with the recent {\it Chandra}
observations reported in Henry (2001), in that 75\% of $z>0.75$
clusters show significant substructure. The fraction of local clusters
and groups with substructure has yet to be fully quantified (see Mohr,
Mathiesen \& Evrard 1999).

\begin{figure}[t]
\centerline{\includegraphics[width=27pc]{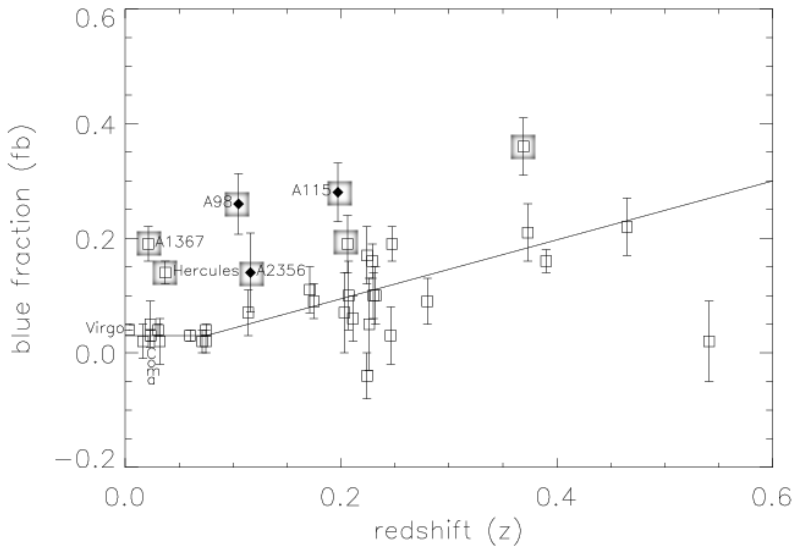}}
\caption[met]{Figure taken from Metevier et al. (2000). This shows the
original Butcher \& Oemler data, as well as new clusters selected by
Metevier et al. We have highlighted (with a gray--shaded box) clusters
with known substructure (or classified as irregular). This is an
incomplete literature search of these clusters, and there may be
more with substructure than presented here.}
\label{met}
\end{figure}

It is natural to ask the question: What effect do such merger events
have on the properties of galaxies within the clusters? Also, could
the change in the merger rate (Figure \ref{merger}) mimic an
evolutionary trend? To partially address these questions, we will use
the Butcher--Oemler (BO) effect as a case study (see also Poggianti
2002 in this volume).

In Figure \ref{met}, we show the original BO effect, {\it i.e.}, the
fraction of blue galaxies in clusters ($f_b$) increasing with redshift
(Butcher \& Oemler 1984). In this figure, we have highlighted clusters
with known substructure in the literature, or that have been
classified as irregular, which suggests these systems have experienced
a recent major interaction (see Metevier et al. 2000; Wang \& Ulmer
1997; Wang, Ulmer, \& Lavery 1997). In fact, all the high $f_b$
clusters in this figure (and thus in the original BO sample, see also
Pimbblet et al. 2002) appear to be merger events; a fact noted by
Butcher and Oemler themselves. Therefore, could the evolutionary trend
of increasing $f_b$ with redshift (seen in Figure \ref{met}) be
primarily due to the increased fraction of disturbed clusters with
redshift, as seen Figure \ref{merger}? For example, the skewness
of the $f_b$ distribution could be increasing with redshift, as more
clusters populate the high $f_b$ tail of the distribution. Kron (1993)
raised similar concerns regarding the possible role of selection
effects in producing the BO effect, while Kauffmann (1995) also proposed
that the BO effect was a natural consequence of hierarchical structure
formation.

\begin{figure}[t]
\centerline{\includegraphics[width=16pc]{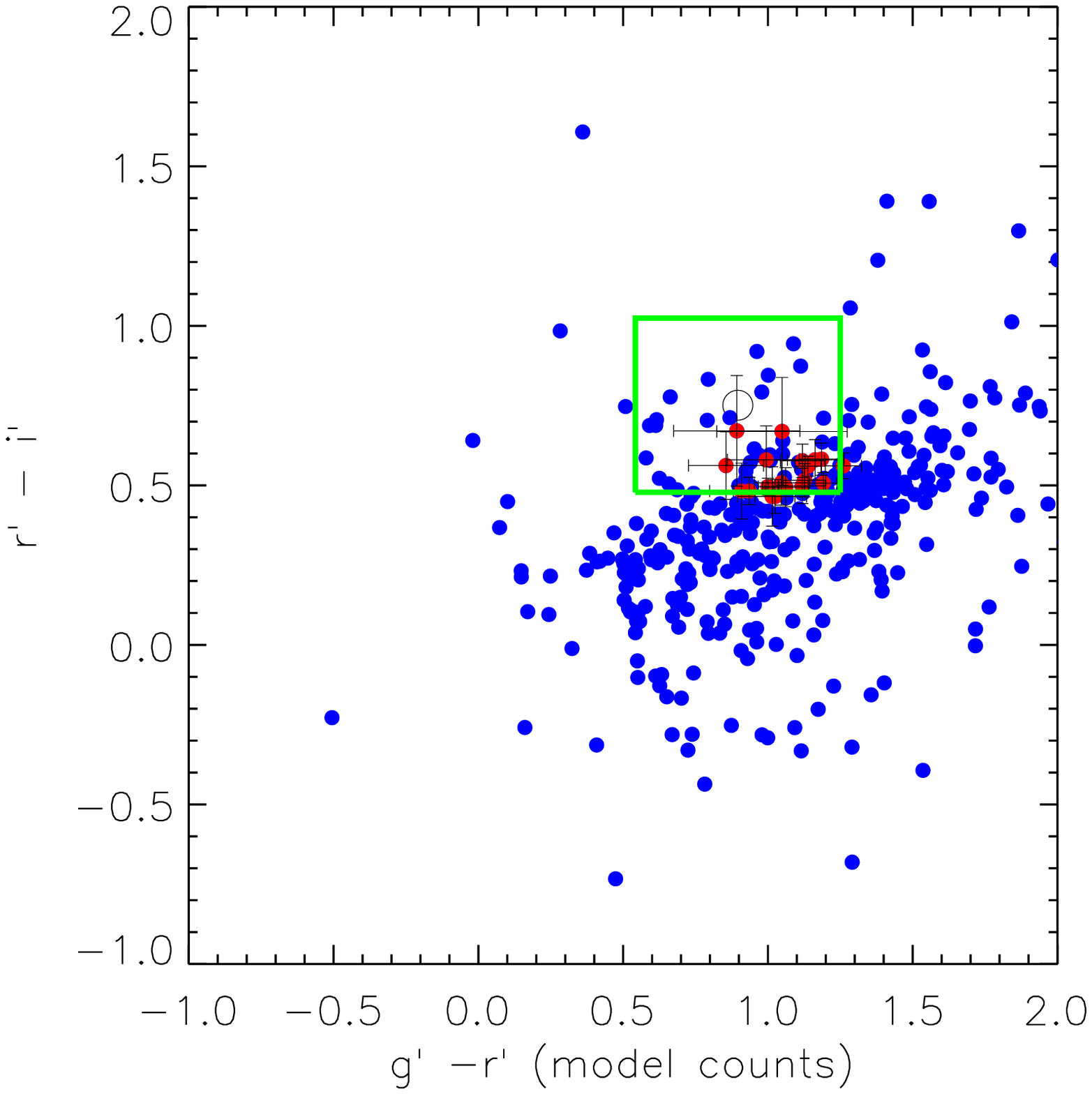}
\includegraphics[width=16pc]{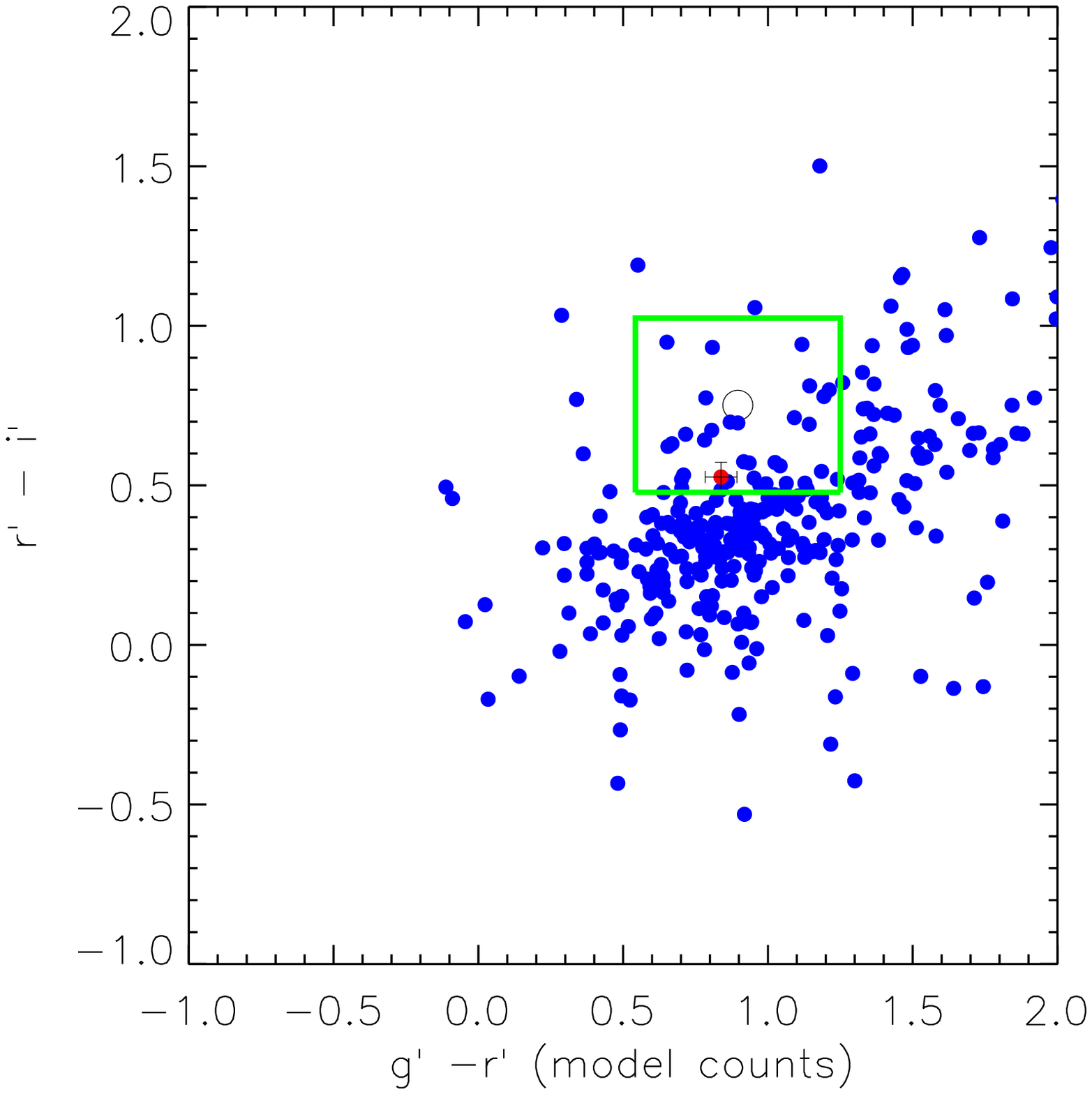}}
\caption{We illustrate here the power of using color--clustering to
find clusters of galaxies. The open circle shown in both plots is the
color of a target galaxy (in a known cluster) compared to the colors
of all neighboring galaxies (blue points) within the spatial part of
the 7--dimensional box placed on this galaxy {\it i.e.}, the part of
the box defined by the RA, DEC and redshift of the target galaxy. Now
we add the color box, which is in green, {\it i.e.}, we show the
projection of the 4--dimensional color box in the g-r, r-i color-color
plane. On the left, we show the known cluster region, and therefore,
the galaxy has many neighbors (shown in red) that are within the
spatial and color box. However, on the right, we show a randomly
chosen field galaxy, which only has one neighbor (in red).  {\it In
summary, there are no projection effects in 7--dimensional space}.}
\label{proj}
\end{figure}

\begin{figure}[t]
\centerline{
\includegraphics[width=30pc]{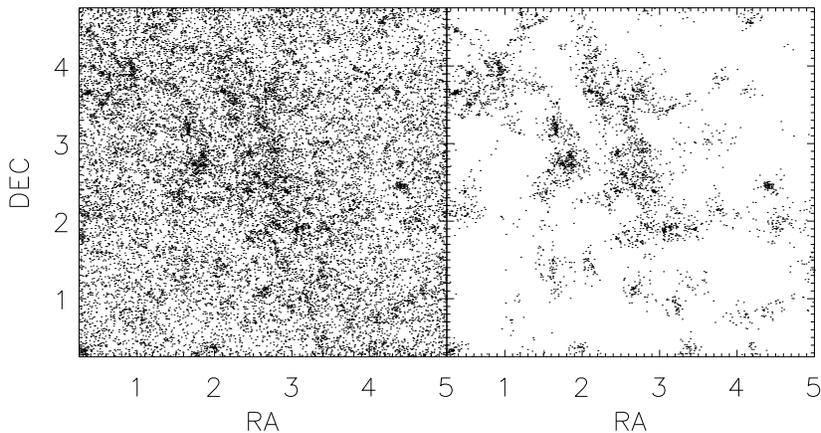}}
\caption{(Right) Projection of all the SDSS data before we
apply the C4 algorithm. (Left) Result of applying the C4 algorithm (only 
$\sim 20\%$ of galaxies remain)}
\label{before}
\end{figure}

To fully test this hypothesis would require a larger, more systematic,
study which must begin with an objective sample of clusters that
possess well-determined physical properties {\it i.e.}, mass,
dynamical state and local environment. Such work has already begun
with X--ray samples of clusters, {\it e.g.}, Wake et al. (in prep)
have systematically studied the blue fraction of galaxies in sixteen,
intermediate redshift clusters, and find no correlation of $f_b$ with
X--ray luminosity or redshift (see also Fairley et al. 2002, Metevier
et al. 2000; Smail et al. 1998; Andreon \& Ettori 1999). Secondly,
Goto et al. (2003a) is studying the BO effect in optical clusters
objectively selected from the Sloan Digital Sky Survey (SDSS)
photometric data (see Goto et al. 2002), while we will discuss below a
new sample of clusters objectively--selected from the SDSS
spectroscopic survey.

We have purposely focused on the BO effect here, at the expense of 
recent morphological and spectroscopic studies of clusters ({\it
e.g.}, CNOC \& MORPHS work). We refer the reader to the review article
of Poggianti (this volume) and Ellingson et al. (2001) for discussion
of this work.  Furthermore, we have ignored other explanations for the
BO effect (both real and systematic).

\section{The SDSS C4 Catalog}

\begin{figure}[t]
\centerline{\includegraphics[width=33pc]{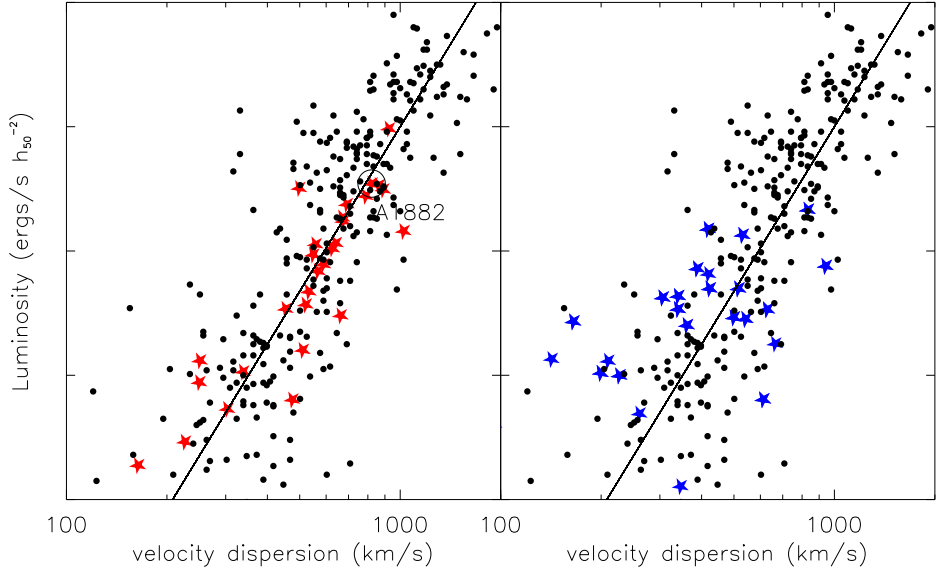}}
\caption{The ${\rm L_x}$--$\sigma_v$ relation for SDSS C4 clusters
selected from the EDR data (star symbols). The black dots are the data
compilation of Mahdavi \& Geller (2001). On the left (in red) are C4
clusters with only one peak in their velocity distribution ({\it
i.e.}, isolated from possible nearby systems), while on the right (in
blue), are C4 clusters with more than one peak in their velocity
distribution ({\it i.e.}, they may have a nearby companion). The line
is the best fit from Mahdavi \& Geller (2001); a slope of 4.4}
\label{lx-sigma}
\end{figure}

The Sloan Digital Sky Survey (SDSS) is underway (see Stoughton et
al. 2002 and {\tt http://www.sdss.org}). As of December 2002, the SDSS has obtained $\sim30\%$ of
the spectra it plans to obtain (corresponding to $\sim300,000$ galaxy
redshifts). When combined with the SDSS multi--color photometry, this
dataset is ideal for objectively finding clusters and groups of
galaxies in three--dimensions.

\begin{figure}[t]
\centerline{\includegraphics[height=22pc,width=27pc]{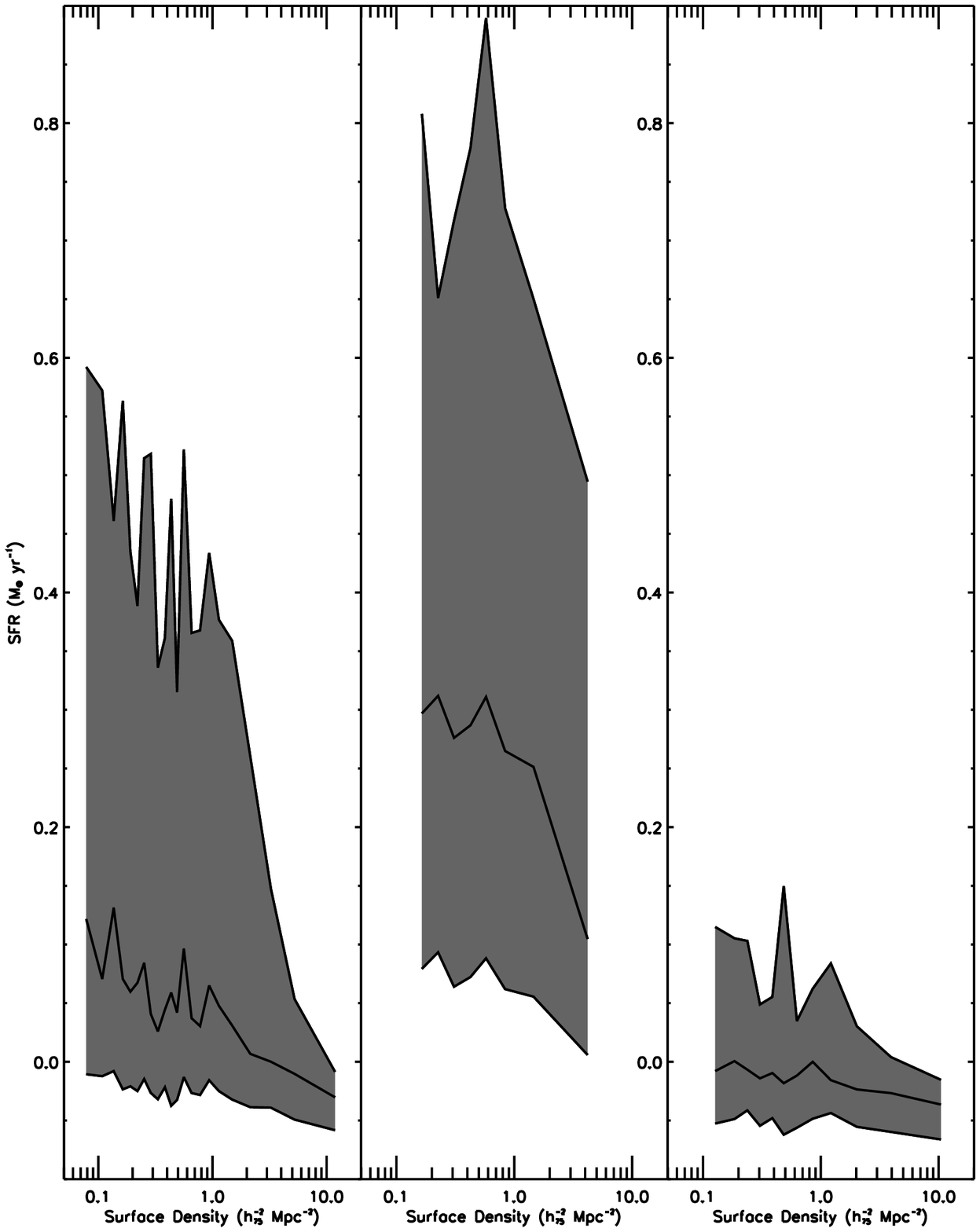}}
\caption{The {\it density--SFR relation} for SDSS galaxies; see Gomez
et al. (2003). In the left-hand panel, we show this relation for all
galaxies regardless of morphology. See the ``break'' in the relation
at a density (x--axis) of $\simeq 1\, {\rm Mpc^{-2}}$. The top of the
shaded area is the $75^{th}$ percentile of the distribution, the
bottom is the $25^{th}$ percentile, while the line is the median. The
middle panel is the same as the left-hand panel, but now for the subset
of galaxies with a concentration index of $C > 0.4$, which is
consistent with late--type (spiral) galaxies. The right--hand panel is
for $C < 0.4$, which are likely to be early--type galaxies. There is
evidence for a density--SFR relation for all galaxy types.}
\label{gomez}
\end{figure}

In Miller et al. (in prep), we outline the C4 algorithm of finding
clusters and groups in the SDSS spectroscopic data (see also Nichol et
al. 2001 \& Gomez et al. 2003). Briefly, the fundamental premise of
the C4 algorithm is that galaxies in the cores of clusters are a
co--evolving population and thus have the same spectral energy
distributions (same colors). Therefore, we perform a search for
clusters of galaxies simultaneously in both redshift--space and
color--space.  This involves counting the number of nearest neighbors
in a 7--dimensional box (3 spatial and 4 color coordinates) placed
around each galaxy in our sample. The size of the box is determined by
the observed errors on the color of the target galaxy, and the
observed scatter in the color--magnitude relation of ellipticals in
clusters. In Miller et al. (in prep), we demonstrate that the clusters
selected using this technique are relatively insensitive to the
details of the algorithm ({\it e.g.}, box size). The power of the C4
approach is that galaxies only sparsely populate the 7--dimensional
space considered here, and therefore, {\bf any} observed clustering in
this high dimensional space is statistically significant, see Figure
\ref{proj}. The C4 algorithm is similar to the work of Gladders \& Yee
(2000) and Goto et al. (2002), but with one key difference; these
other techniques model the expected colors of cluster galaxies, as a
function of redshift, while the C4 algorithm is model--independent and
only requires a clustering in color--space, rather than a specific
color.

Once we have counted the number of nearest neighbors, we used the
False Discovery Rate (FDR; Miller et al. 2001) to determine which
galaxies have an unusual number of nearest neighbors in the
7--dimensional box, {\it i.e.}, live in clusters. This thresholding
scheme removes 80\% of all galaxies, preferentially those in
low--density regions (see Figure \ref{before}). Next, these remaining
galaxies are clustered using a density estimator and cluster
candidates selected. Once we have a list of candidates, we then
measure a variety of physical parameters, including the velocity
dispersion, total optical luminosity and the existence of any nearby
companion. We also use the ROSAT All-Sky Survey data to estimate the
X--ray luminosity of each cluster. Details of the C4 algorithm,
including extensive tests of the completeness and purity of the
catalog, can be found in Miller et al. (in prep).

In Figure \ref{lx-sigma}, we present first results from running the C4
algorithm on the SDSS Early Data Release (EDR; Stoughton et
al. 2002). We show the relation between X--ray luminosity and velocity
dispersion for EDR clusters, as a function of their morphology. The
cluster morphology is based on the number of peaks detected in the
velocity distribution of each cluster. As one can see, isolated
clusters have a tight ${\rm L_x}$--$\sigma_v$ relation; about a factor
of 5 tighter than previous observations (Mahdavi \& Geller 2001) or
EDR clusters with a complex morphology (see also Smith et al. 2002).

We have begun studying galaxy evolution in SDSS C4 clusters, {\it
e.g.}, in Figure \ref{gomez}, we show the {\it density--star formation
rate (SFR) relation} recently discussed in Gomez et al. (2003). This
work shows a clear ``break'' in the density--SFR relation of galaxies
at a density of $\simeq 1 {\rm Mpc^{-2}}$ (see also Kodama et
al. 2001; Lewis et al. 2002). At densities greater than this (which
corresponds to cluster regions within $\sim2$ virial radii), the SFR
of galaxies decreases rapidly. Below this critical density, the SFR of
galaxies appears to be uncorrelated with density. In Gomez et
al. (2003), we attempted to determine if the density--SFR relation is
degenerate with the density--morphology relation (see Dressler et
al. 1997). As shown in Figure \ref{gomez}, there is some evidence that
these two relations are independent, and all morphological types
experience a density--SFR relation (see Balogh et al. 1998; Poggianti
et al. 1999). This will require further study, but this work does
illustrate the power of the SDSS data and, in the near future, we will
investigate the density--SFR--morphology relation over a range of C4
clusters. We plan to study the BO effect in C4 clusters, as a function
of cluster morphology and luminosity, for spectroscopically confirmed
galaxy members.

\section{Coma Cluster and Post--Starburst Galaxies}

The Coma Cluster is under--going a major merger event with a satellite
group centered on NGC4839 (mass ratio of 3:1; see Neumann et al
2001). Due to the low redshift of Coma, it is ideal for studying the
interplay of such an event with the properties of galaxies in the main
cluster and the group. In fact, the most striking evidence yet for any
interplay between a cluster merger and galaxy evolution, was the
apparent discovery of an excess of post--starburst galaxies (galaxies
with the signature of a recent burst of star--formation) along the
filament of galaxies joining the core of Coma with the NGC4839 group
(see Caldwell \& Rose 1997; Burns et al. 1994).  However, Castander et
al. (2001) recently presented the first hour of extragalactic
observations of the SDSS spectroscopic survey and obtained spectra for
196 Coma galaxies out to the virial radius of Coma (1.5
degrees). These observations covered the southern part of the cluster,
including the NGC4839 group. Castander et al. discovered several
H$\delta$--strong galaxies, indicative of post--starburst (or k+a)
galaxies, but found no evidence that galaxy star--formation (both past
and present) was at all correlated with the existence of a merging
group; see Figure 3 in Castander et al. (2001), as well as Carter et
al. (2002).  Alternatively, the SDSS Coma spectral data are consistent
with simple galaxy in--fall as in Gomez et al. (2003), {\it i.e.}, as
galaxies in--fall into a cluster, their star--formation is slowly
truncated (see Figure \ref{gomez}). The lack of an obvious connection
between the merger in Coma and the galaxy populations may be because
the NGC4839 group is still in--falling, and has yet to pass through the
core of Coma (see Neumann et al. 2001).

We take this opportunity to briefly mention a new sample of
H$\delta$--selected galaxies ({\it i.e.}, post--starburst, k+a) by
Goto et al. (2003b). They have found 3340 such galaxies in the SDSS
spectroscopic sample, which is two orders of magnitude greater than
any other local sample of H$\delta$ galaxies. Due to page constraints,
we will presents results from this work elsewhere.

\section{Summary}

Using two case studies (Coma and the BO effect), we have examined the
interplay between cluster and galaxy evolution. It is clear from this
brief study that both major mergers and galaxy in--fall are important
mechanisms for explaining galaxy evolution in dense environments. As
one goes to higher redshifts, the relative importance of these two
mechanisms may change and therefore, it is imperative that we are
careful about the selection of clusters used in such studies.  Larger,
more homogeneous samples of clusters ({\it e.g.} from XMM and SDSS),
with quantified properties (mass, luminosity, morphology) are
important. Such samples will allow us to robustly investigate the
relation between cluster parameters and the galaxy populations in
those clusters\footnote{Kauffmann (1995) notes that one must account
for cluster evolution when comparing low redshift clusters with higher
redshift systems, as the density of massive clusters is evolving with
redshift.}. 

\acknowledgements We thank our collaborators; Michael Balogh, Ann
Zabludoff, Percy Gomez, Gus Evrard, Risa Wechsler, Tim
Mckay and our SDSS colleagues. We thank Ian Smail, Michael Balogh and
Kathy Romer for reading an earlier draft of this review. RCN thanks
Catarina Lobo and Pedro Viana for their hospitality which made his
stay in Portugal so much fun. Finally, RCN acknowledge partial funding
from the conveners of the "Galaxy evolution" workshop through project
ref. PESO/PRO/15130/1999 from FCT/ Portugal.

\end{article}
\end{document}